# Atomic Electronic States: the *L-S* and *j-j* Coupling Schemes and Their Correlation

**Wai-Kee Li**

Department of Chemistry, The Chinese University of Hong Kong, Shatin, N.T., Hong Kong
wkli@cuhk.edu.hk

**S. M. Blinder**

Department of Chemistry, The University of Michigan, Ann Arbor, MI 48109-1055 and Wolfram Research Inc., Champaign, IL 61820-7237
sblinder@umich.edu; sblinder@wolfram.com

**Abstract**

In the first part of this paper, we review the assumption of the *L-S* coupling scheme, with which we derive the electronic states arising from a given atomic configuration. Then, with the aid of the spectral data of Group 15 elements (configuration $np^3$, $n = 2 - 6$), it becomes clear that the assumption of the *L-S* coupling scheme is no longer valid as we go farther and farther down the Periodic Table. In the second part, we introduce the *j-j* coupling scheme, which is seldom covered in standard inorganic chemistry texts, and contrast the assumptions of the two schemes. Next, we use two worked examples to demonstrate the derivation of electronic states with the *j-j* coupling scheme. Finally, the correlation between the states derived by *L-S* and *j-j* schemes is pictorially shown. It is believed a student, by also studying *j-j* coupling schemes (by no means a difficult task) along with the *L-S* scheme, will gain a better understanding of the concept of atomic electronic states.

**Keywords:** Atomic electronic states; Term symbols; Russell-Saunders or *L-S* coupling scheme; *j-j* coupling scheme and its application; Correlation of the states obtained with *L-S* and *j-j* coupling schemes.

**Introduction**

The derivation of the electronic states arising from a given atomic configuration is by now a conventional topic in a junior- or senior-level inorganic chemistry course (*1-4*). The method usually employed is the well-known *L-S*, or Russell-Saunders, coupling scheme. While the method is relatively straightforward and students have no difficulty in mastering the technique, the physical basis of the scheme is often neglected in the textbook presentation. As a result, students are not familiar with the basic assumption behind the *L-S* coupling scheme and they do not know that assumption starts to break down as we descend the Periodic Table. Indeed, for elements of the fifth and sixth periods, an alternative method, the *j-j* coupling scheme, should be employed instead. It is regrettable that the *j-j* coupling scheme is seldom covered in an undergraduate course or in standard inorganic chemistry texts. In a couple of books (*3-4*) that this scheme is mentioned, the coverage is limited to about three lines of text. Hence, not much can be said other than: "The coupling scheme most appropriate to heavy atoms (that is, atoms of the 4*d* and 5*d* series of elements) is called *j-j* coupling, but we shall not consider it further(*3*)." Or, "Russell-Saunders coupling is not valid for all elements (especially those with high atomic numbers). In an alternative method of coupling, $\ell$ and *s* for all the individual electrons are first combined to give *j*, and the individual *j* values are combined in a *j-j* coupling scheme(*4*)."

In this note, we will first give spectral evidence to show that the *L-S* coupling scheme is not valid for all elements in the Periodic Table and, for very heavy atoms, *j-j* coupling should be used instead. Next, we will use two worked examples to demonstrate the derivation of the electronic states arising from a given configuration with the *j-j* coupling scheme. We will then illustrate pictorially how the two sets of energy levels, derived with *L-S* and *j-j* coupling schemes, correlate with each other. The treatment given here is totally non-mathematical. There are no equations in the discussion and essentially all mathematical expressions within the text are arithmetical.

**The *L-S* coupling scheme: Its assumption and limitation**

When we derive the electronic states arising from a given atomic configuration, we take advantage of the simplification that all the complete shells of inner core electrons contribute no net orbital and spin angular momenta (excepting the consideration of X-ray and ESCA spectra). Apart from the attraction between the nucleus and electrons, there remain two principal contributions to the energy in an atomic system: (1) inter-electronic repulsion, which is responsible for the energy separation between the terms arising from a given configuration, and (2) spin-orbit interaction, which gives rise to the energy intervals between the components within a given term. In the *L-S* coupling scheme, it is assumed that electronic repulsion is much greater than spin-orbit interaction. In the terminology of quantum mechanics, the spin-orbit interaction in this scheme is treated as a perturbation.

The *L-S* coupling scheme involves the orbital angular momentum quantum number *L* and the spin angular momentum quantum number *S*. They define an electronic term conventionally designated by a term symbol written in the form $^{2S+1}L_J$. The addition of *L* and *S* gives the total angular momentum quantum number *J*, which has the allowed values given by the sequence: $L + S, L + S - 1, \ldots, |L - S|$. This results in a multiplicity of $2S + 1$ for a given orbital angular momentum *L* (except when $L = 0$) which is written as a subscript preceding the code symbol *S, P, D, F, …* for $L = 0, 1, 2, 3, \ldots$. Beware of the possible confusion between the code symbol *S* for $L = 0$ and the total spin angular momentum quantum number, which is also designated as *S*. Spin-orbit interaction between the total orbital and spin angular momenta give rise to different energies for each allowed value of *J*. For the lighter elements, these energy separations are relatively small.

When atomic states are accurately represented by Russell-Saunders coupling, the energy ordering of different terms arising from a given electron configuration follow Hund's rules(*5*), which can be summarized as follows: (1) higher multiplicities have lower energies and (2) for terms of the same multiplicity, larger *L* values have lower energies. For example, the valence electron configuration of $2p^2$ for the carbon atom gives rise to three terms: $^3P < {}^1P < {}^1D$. Spin-orbit interaction between *L* and *S* then causes relatively small intervals of each term for different possible values of *J*. For example, the lowest-energy term for the carbon atom can have three possible *J* values, with the energy-level ordering: $^3P_0 < {}^3P_1 < {}^3P_2$. It is clear, with this scheme, each electronic state is defined by quantum numbers *L*, *S*, and *J*.

Let us now explore the quantitative validity of the Russell-Saunders coupling scheme. How well does it hold and when does it begin to break down? These questions can be answered by referring to some atomic spectral data. For example, we have taken from the literature(*6*) the atomic energy levels arising from configurations $np^3$, $n = 2 - 6$, of Group 15 elements. Applying the *L-S* coupling scheme to the $p^3$ configuration, we get three spectroscopic terms(*7*), namely $^4S$, $^2D$, and $^2P$, which give rise to five energy levels,

$^4S_{1½}$ (ground state), $^2D_{1½}$, $^2D_{2½}$, $^2P_{½}$, and $^2P_{1½}$. The energy values of these five levels for all Group 15 elements are listed in Table 1. Also tabulated there are the "weighted" term values for $^2D$ and $^2P$. The energy of $^2D$(weighted) is simply (6 × $^2D_{2½}$ + 4 × $^2D_{1½}$)/10, where the numerical factors "6" and "4" refer to the number of components (2$J$ + 1) of the states $^2D_{2½}$ and $^2D_{1½}$, respectively. The energy of $^2P$(weighted) can be calculated in a similar manner. These "weighted" term values provide us with the approximate energy of the term, before spin-orbit interaction is "turned on."

So how good is the assumption of the *L-S* coupling scheme that electronic repulsion is much, much larger than the spin-orbit interaction? If we look at the data for the nitrogen atom, we find the two energy gaps between the three terms (which give a measure of the electronic repulsion) are of the order of $10^4$ cm$^{-1}$. On the other hand, the intervals between the two states arising from the same term (which accounts for the spin-orbit interaction within the term) are only a few cm$^{-1}$. Based on these results, the aforementioned assumption is clearly justified. For the other four atoms within the same Group, the two energy gaps among the three terms remain to be about $10^4$ cm$^{-1}$. But the spin-orbit interaction increases by about an order of magnitude as we go down the Periodic Table: $10^0$ to $10^1$ to $10^2$ to $10^3$ to $10^4$ cm$^{-1}$ as we go from N to P to As to Sb to Bi. Also, for the lighter elements, above the ground state $^4S_{2½}$, we can clearly see two sets of "doublet" levels, with very small to relatively small intervals between the two levels. But such a description no longer exists as we go to the sixth row of the Periodic Table. Indeed, for Bi, the $^2P_{1½}$ level is farther away from $^2P_{½}$ (both of these two states originating from the same term) than from $^2D_{1½}$ (a state arising from a different term)! This picture is certainly not envisaged by the *L-S* coupling scheme. Clearly, for Bi, the *L-S* coupling scheme is not satisfactory.

Before leaving this topic, it should be noted that the breakdown of the *L-S* coupling scheme for the heavier elements is by no means merely an empirical trait. Rather, it has a theoretical basis. The spin-orbit interaction energy in a hydrogen-like atom can be calculated exactly(*8-10*) and it is proportional to the ratio of $Z^4/n^3$, where $Z$ is the atomic number and $n$ the principal quantum number of the outer electrons. It is expected that this proportionality holds approximately for other atoms in the Periodic Table.

**An alternative scheme: *j-j* coupling**

In the *j-j* coupling scheme, in contrast to the *L-S* method, we assume the spin-orbit interaction is much larger than the electronic repulsion. However, the spin-orbit interactions in these two methods are somewhat different. In the *L-S* coupling scheme, *for the whole system*, the orbital angular momentum vector (with quantum number $L$) interacts with the spin angular momentum vector (with quantum number $S$), yielding the total angular momentum vector (with quantum number $J$), and these three quantum numbers define the electronic states of an atom. On the other hand, in the *j-j* coupling scheme, each electron's $\ell$ and $s$ first combined to give $j$, i.e., $j = \ell - \frac{1}{2}$ and $\ell + \frac{1}{2}$. Then the $j$ values of individual electrons are combined to yield total angular momentum quantum number $J$. Here we follow the convention of using the upper-case letters to denote the properties of the whole atom and lower-case letters for the properties of individual electrons. Clearly, in the *j-j* scheme, $L$ and $S$ are no longer "good" quantum numbers and only $J$ retains its validity. That is to say, with the *j-j* coupling scheme, each atomic electronic state is defined by $J$ alone.

In this section, we will use two examples to illustrate the correlation between the *L-S* and *j-j* coupling schemes. The first configuration to be treated is $s^1 f^1$, a system with two non-equivalent electrons. In the *L-S* scheme, we have $\ell_1 = 0$ and $\ell_2 = 3$ and hence $L = \ell_1 + \ell_2, \ell_1 + \ell_2 - 1, \ldots, |\ell_1 - \ell_2| = 3$. For the spin part, $s_1 = s_2 = \frac{1}{2}$ and hence $S = s_1 + s_2, s_1 + s_2 - 1, \ldots, |s_1 - s_2| = 1$ and $0$. So, for the $s^1 f^1$ configuration, there are two terms, $^3F$ and $^1F$. Turning on spin-orbit interaction gives four states: $^3F_2$, $^3F_3$, $^3F_4$, and $^1F_3$. On the other hand, in the *j-j* scheme, we have $\ell_1 = 0$ and $s_1 = \frac{1}{2}$, hence $j_1 = \ell_1 + s_1, \ell_1 + s_1 - 1, \ldots, |\ell_1 - s_1| = \frac{1}{2}$. Also, $\ell_2 = 3$ and $s_2 = \frac{1}{2}$, hence, similarly, $j_2 = 2\frac{1}{2}$ and $3\frac{1}{2}$. When we couple $j_1 = \frac{1}{2}$ with $j_2 = 2\frac{1}{2}$, we get $J = j_1 + j_2, j_1 + j_2 - 1, \ldots, |j_1 - j_2| = 2$ and $3$. When we combine $j_1 = \frac{1}{2}$ with $j_2 = 3\frac{1}{2}$, we get $J = 3$ and $4$. In other words, both methods come up with four energy levels, with $J$ values of 2, 3 (twice), and 4. The correlation of these two sets of electronic states is shown schematically in Figure 1.

The second example is configuration $p^3$, for which the *L-S* results were discussed in the previous section. This system consists of three equivalent electrons, hence we need to take account of the Exclusion Principle. Within the *j-j* scheme, the Exclusion Principle implies that "no two electrons can have the same set of quantum numbers $(n, \ell, j, m_j)$." For configuration $p^3$, we have $\ell_1 = \ell_2 = \ell_3 = 1$, $s_1 = s_2 = s_3 = \frac{1}{2}$, and $j_1, j_2, j_3 = \frac{1}{2}, 1\frac{1}{2}$. When $j_1, j_2, j_3 = \frac{1}{2}$, $(m_j)_1$, $(m_j)_2$, and $(m_j)_3$ can be $\frac{1}{2}$ or $-\frac{1}{2}$. When $j_1, j_2, j_3 = 1\frac{1}{2}$, $(m_j)_1$, $(m_j)_2$, and $(m_j)_3$ can be 1½, ½, –½ or –1½.  Consistent with the Exclusion Principle, we arrive at 20 microstates, each of which is described by the combination [$j_1, j_2, j_3, (m_j)_1, (m_j)_2, (m_j)_3$]. These 20 microstates are listed in Table 2.  Examining this table, we find: (1) aggregate ($j_1$ = ½, $j_2$ = ½, $j_3$ = ½) does not lead to any [$j_1, j_2, j_3, (m_j)_1, (m_j)_2, (m_j)_3$] combination that satisfies the Exclusion Principle; (2) aggregates ($j_1$ = ½, $j_2$ = ½, $j_3$ = 1½) and ($j_1$ = 1½, $j_2$ = 1½, $j_3$ = 1½) yield one state each, both with $J$ = 1½; (3) the ($j_1$ = ½, $j_2$ = 1½, $j_3$ = 1½) aggregate yields three states, with $J$ = ½, 1½, and 2½.  In summary, the $j$-$j$ scheme comes up with five energy levels, with $J$ = ½, 1½ (thrice), and 2½.  Recalling from the previous section, the $L$-$S$ scheme also leads to five energy levels: $^4S_{1½}$, $^2D_{1½}$, $^2D_{2½}$, $^2P_{½}$, and $^2P_{1½}$. The $J$ values of these two sets of states are identical, as should be the case.  The correlation of the two sets of electronic states arising from the $p^3$ configuration is schematically shown in Figure 2.  Finally, it is important to note that in Table 2 the electronic labels 1, 2, and 3 are included merely for the sake of convenience; we need to bear in mind that all electrons are indistinguishable from each other.

If we may be allowed a humorous side note, some physics undergraduates at Cambridge University once invented a fictitious professor called J. J. Coupling, an obvious spoof on J. J. Thomson, who discovered the electron.

**Summary**

This note consists of two parts.  In the first part, we discuss the basic assumption of the $L$-$S$ coupling scheme, with which we derive the electronic states of a given electronic configuration.  In addition, we use the spectral data of Group 15 elements (configuration $np^3$, $n$ = 2 – 6), to show that the assumption of the $L$-$S$ coupling scheme becomes less satisfactory as we go further and further down the Periodic Table.  In the second part, we introduce the $j$-$j$ coupling scheme, which is seldom covered in standard inorganic chemistry texts.  We then use two worked examples, one involving non-equivalent electrons and the other containing equivalent electrons, to illustrate how to apply the $j$-$j$ scheme to derive the electronic states arising from an atomic configuration. Finally, a correlation between the states derived by $L$-$S$ and $j$-$j$ schemes is shown pictorially.  By studying both the $j$-$j$ and the $L$-$S$ coupling schemes, a student will gain a deeper, as well as a broader, perception of the concept of atomic electronic states.

Table 1. Spectral data (in cm$^{-1}$) for the electronic states
arising from configurations $np^3$, $n$ = 2 - 6

|   | $^4S_{1½}$ | $^2D_{1½}$ | $^2D_{2½}$ | $^2D$(weighted)† | $^2P_{½}$ | $^2P_{1½}$ | $^2P$(weighted)† |
|---|---|---|---|---|---|---|---|
| N  | 0.0 | 19231    | 19223    | 19226    | 28840*   | 28840*   | 28840    |
| P  | 0.0 | 11361.7  | 11376.5  | 11367.6  | 18722.4  | 18748.1  | 18731.0  |
| As | 0.0 | 10592.5  | 10914.6  | 10721.3  | 18186.1  | 18647.5  | 18339.9  |
| Sb | 0.0 | 8512.1   | 9854.1   | 9048.9   | 16395.6  | 18464.5  | 17085.2  |
| Bi | 0.0 | 11419.03 | 15437.66 | 13026.48 | 21661.00 | 33164.84 | 25495.61 |

† $^2D$(weighted) = (6 × $^2D_{2½}$ + 4 × $^2D_{1½}$)/10; $^2P$(weighted) can be obtained analogously.

* These two lines have not been resolved.

Table 2.  The microstates for configuration $np^3$ employing the *j-j* coupling scheme

| $j_1$ | $j_2$ | $j_3$ | $(m_j)_1$ | $(m_j)_2$ | $(m_j)_3$ | $M_J$ | $J$ |
|---|---|---|---|---|---|---|---|
| ½ | ½ | 1½ | ½ | -½ | 1½ | 1½ | |
| | | | ½ | -½ | ½ | ½ | 1½ |
| | | | ½ | -½ | -½ | -½ | |
| | | | ½ | -½ | -1½ | -1½ | |
| ½ | 1½ | 1½ | ½ | 1½ | ½ | 2½ | |
| | | | ½ | 1½ | -½ | 1½ | |
| | | | ½ | 1½ | -1½ | ½ | |
| | | | ½ | ½ | -½ | ½ | |
| | | | ½ | ½ | -1½ | -½ | |
| | | | ½ | -½ | -1½ | -1½ | |
| | | | -½ | 1½ | ½ | 1½ | ½, 1½, 2½ |
| | | | -½ | 1½ | -½ | ½ | |
| | | | -½ | 1½ | -1½ | -½ | |
| | | | -½ | ½ | -½ | -½ | |
| | | | -½ | ½ | -1½ | -1½ | |
| | | | -½ | -½ | -1½ | -2½ | |
| 1½ | 1½ | 1½ | 1½ | ½ | -½ | 1½ | |
| | | | 1½ | ½ | -1½ | ½ | 1½ |
| | | | 1½ | -½ | -1½ | -½ | |
| | | | ½ | -½ | -1½ | -1½ | |

Figure 1. Correlation of electronic states arising from the $s^1f^1$ configuration applying *L-S* and *j-j* coupling schemes.

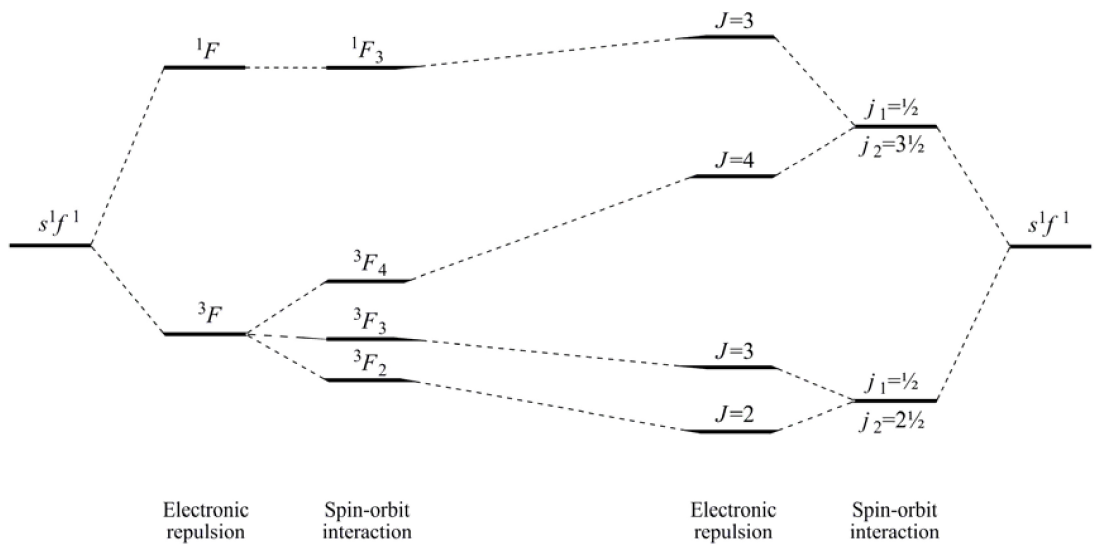

Figure 2. Correlation of electronic states arising from the $p^3$ configuration applying *L-S* and *j-j* coupling schemes.

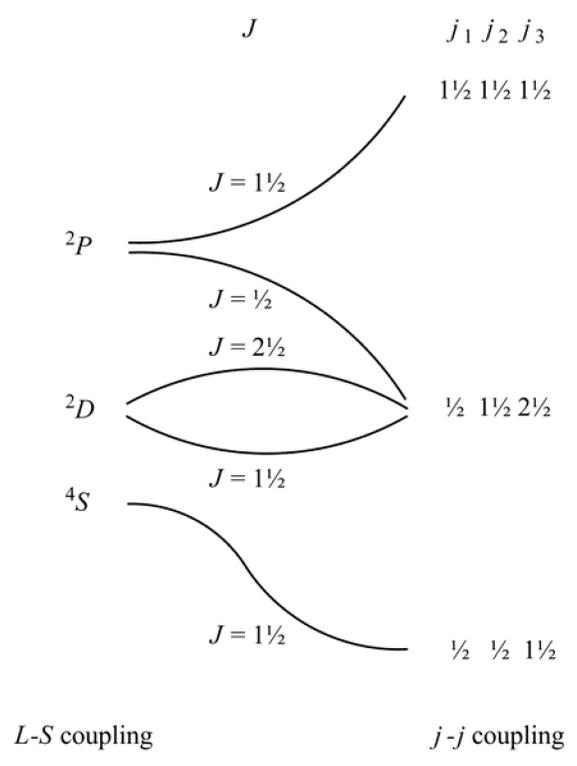